\documentclass[conference]{IEEEtran}

\makeatletter

\def\ps@IEEEtitlepagestyle{%
  \def\@oddfoot{\mycopyrightnotice}%
  \def\@evenfoot{}%
}
\def\mycopyrightnotice{%
  {\footnotesize XXX-X-XXXX-XXXX-X/XX/\$XX.00~\copyright~20XX IEEE\hfill}
  \gdef\mycopyrightnotice{}
}

\usepackage{blindtext}
\usepackage{eso-pic}
\IEEEoverridecommandlockouts
\usepackage{cite}
\usepackage{amsmath,amssymb,amsfonts}
\usepackage{algorithmic}
\usepackage{graphicx}
\usepackage{textcomp}
\usepackage{xcolor}
\usepackage{graphicx}
\usepackage{tabularx}
\usepackage{array}
\usepackage{dirtytalk}
\usepackage[export]{adjustbox}
\usepackage{multirow}
\usepackage{graphicx}
\usepackage{float}
\usepackage[justification=justified, format=plain]{caption}
\usepackage{subcaption}
\usepackage{hhline}
\newcolumntype{C}[1]{>{\centering\let\newline\\\arraybackslash\hspace{2pt}}m{#1}}
\def\BibTeX{{\rm B\kern-.05em{\sc i\kern-.025em b}\kern-.08em
    T\kern-.1667em\lower.7ex\hbox{E}\kern-.125emX}}
    
\usepackage{eso-pic}
\newcommand\AtPageUpperMyright[1]{\AtPageUpperLeft{%
 \put(\LenToUnit{0.17\paperwidth},\LenToUnit{-2cm}){%
     \parbox{0.9\textwidth}{\raggedleft\fontsize{8}{11}\selectfont #1}}%
 }}%
\newcommand{\conf}[1]{%
\AddToShipoutPictureBG*{%
\AtPageUpperMyright{#1}
}
}

\begin{document}
\title{\vspace*{1cm} UNet Based Pipeline for Lung Segmentation from Chest X-Ray Images
}

\author{\IEEEauthorblockN{Shashank Shekhar\textsuperscript{1}}
\IEEEauthorblockA{\textit{Department of Computer Engineering} \\
\textit{New York University}\\
New York, 11201, United States \\
shashank.shekhar@nyu.edu}
\\
\IEEEauthorblockN{H Srikanth Kamath\textsuperscript{2}}
\IEEEauthorblockA{\textit{Department of Electronics and Communication Engineering} \\
\textit{Manipal Institute of Technology, Manipal Academy of Higher Education}\\
Manipal, 576104, India \\
srikanth.kamath@manipal.edu}
\\
\IEEEauthorblockN{Ritika Nandi\textsuperscript{3}}
\IEEEauthorblockA{\textit{Department of Applied Analytics} \\
\textit{School of professional Studies, Columbia University}\\
New York, 10025, United States \\
ritika.nandi@columbia.edu}
}

\maketitle
\conf{\textit{  Proc. of the International Conference on Electrical, Computer and Energy Technologies (ICECET 2022) \\ 
20-22 June 2022, Prague-Czech Republic}}
\begin{abstract}
Biomedical image segmentation is one of the fastest growing fields which has seen extensive automation through the use of Artificial Intelligence. This has enabled widespread adoption of accurate techniques to expedite the screening and diagnostic processes which would otherwise take several days to finalize. In this paper, we present an end-to-end pipeline to segment lungs from chest X-ray images, training the neural network model on the Japanese Society of Radiological Technology (JSRT)  dataset, using UNet to enable faster processing of initial screening for various lung disorders. The pipeline developed can be readily used by medical centers with just the provision of X-Ray images as input. The model will perform the preprocessing, and provide a segmented image as the final output. It is expected that this will drastically reduce the manual effort involved and lead to greater accessibility in resource-constrained locations.
\end{abstract}



\section{Introduction}
Time is of essence in the medical field, where a patient’s situation might change drastically in a matter of days. In light of the extremely time-sensitive nature of the field, any method to reduce delays needs to be analyzed thoroughly and implemented wherever possible. Artificial Intelligence coupled with large scale data has enabled the proliferation of technology in the medical domain \cite{sicario}, rapidly replacing human effort with nearly the same or even increased accuracy. Various benchmarks have been established for image segmentation \cite{yoyo,oyoy}, object detection \cite{pewpew}, and identification \cite{hoehoe}; and new and improved deep learning techniques have been found to consistently perform better than humans at such tasks \cite{tuntun}. This is because the architecture of neural networks makes it easy for them to downsample the images, thereby extracting localized features, which can then be combined to effect highly accurate results \cite{gurr}, in a fraction of the time a normal human would take.
Another factor apart from time is cost, and in a field already stretched thin in terms of resources, even marginal errors might lead to a massive increase in costs for patients. This also discourages a vast section of the people from seeking medical advice in the first place, owing to the time-consuming and expensive nature of the process. A reduction in processing time means a reduction in cost for patients too, since one system can efficiently perform the task of multiple technicians, with its accuracy increasing with the number of images it processes, utilizing the concept of backpropagation \cite{panda}.
It is thus logical to infer that efforts need to be made to optimize the diagnostic pipeline, and state-of-the-art techniques developed in this regard would have far-reaching effects on the general well-being of the society. In order to proceed in this direction, this paper proposes a UNet architecture \cite{potato} for lung segmentation from chest X-ray images, thus shortening the otherwise manual process, with a considerable spike in accuracy over other image processing approaches.

\section{Problem Statement}
An essential part of strategizing towards solving any problem is the formulation of a comprehensive and lucid statement describing the problem at hand, the data available, edge cases if any, and the expected benchmarks.
The current problem statement can be described as \emph{Given an X-ray image of the chest, outline and segment out the lungs, using any medical dataset for training and validation purposes, taking care that lungs may sometimes be obscured or incompletely captured in the X-Rays, accounting for this discrepancy through suitable measures and providing an output with a satisfactory DICE score} \cite{pot}.

\begin{figure*}
\begin{center}
\includegraphics[width=30pc]{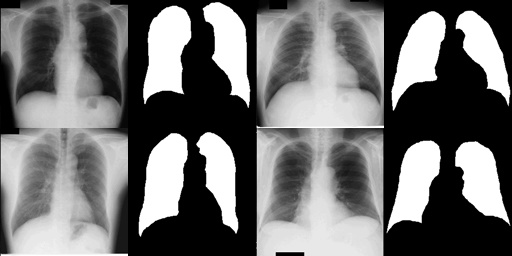}
\end{center}
\caption{\label{fig1}Sample of the JSRT dataset \cite{hotcum} with X-ray images and \\ corresponding lung masks}
\end{figure*}

\section{Dataset}
The dataset used was the Standard Digital Image Database of chest radiographs with and without a lung nodule which was developed as part of the research activities of the Academic Committee of the Japanese Society of Radiological Technology (JSRT) in 1998. \say{\emph{The database includes 154 conventional chest radiographs with a lung nodule (100 malignant and 54 benign nodules) and 93 radiographs without a nodule which were digitized by a laser digitizer with a 2048x2048 matrix size (0.175-mm pixels) and a 12-bit gray scale (no header, big-endian raw data). The database also includes additional information such as; patient age, gender, diagnosis (malignant or benign), X and Y coordinates of nodule, simple diagram of nodule location}} \cite{hotcum}.

The JSRT database has been widely used in research projects of varying scales worldwide for diverse applications like image processing, image compression \cite{jizz}, evaluation of image display, picture archiving and communication system (PACS) \cite{ballz}, and computer-aided diagnosis algorithms \cite{hi}. 

It was deemed suitable for our work since the dataset was widely accepted and benchmarked, and had been refined through years of preprocessing by various researchers. Since the dataset consisted of a relatively small number of data points, it was also suitable for low-resource intensive applications leading to greater accessibility. The Figure \ref{fig1} provides a small sample of the dataset for reference as to what the input would ideally look like.

\begin{figure*}
	\begin{subfigure}{0.48\textwidth}
    	\includegraphics[width=\linewidth, , height=10.5cm, frame]{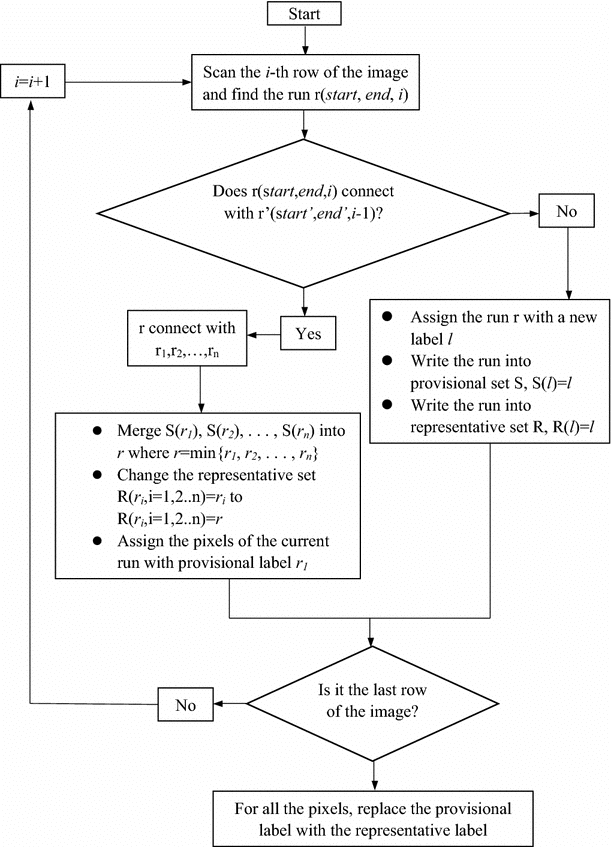}
		\caption{Connected Component Analysis (CCA)}
		\label{fig21}
	\end{subfigure}
	\begin{subfigure}{0.48\textwidth}
		\includegraphics[width=\linewidth, height=10.5cm, frame]{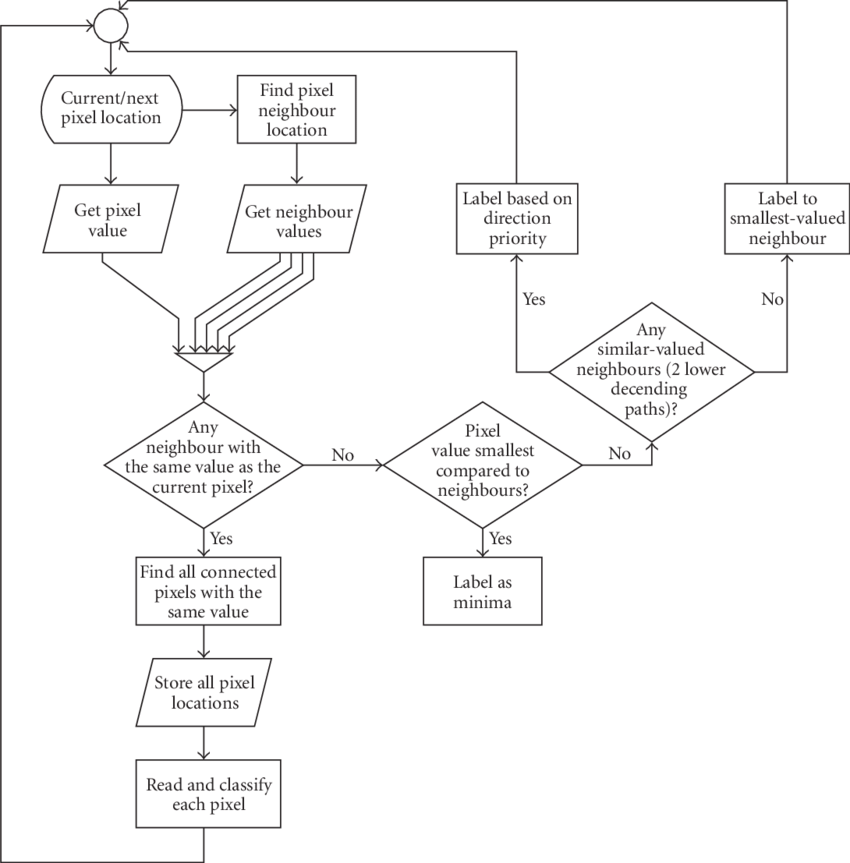}
		\caption{Watershed Algorithm}
		\label{fig22}
	\end{subfigure}
	\caption{Flowcharts of the algorithmic approaches}
	\label{fig2}
\end{figure*}

\begin{figure*}
\begin{center}
\includegraphics[width=30pc]{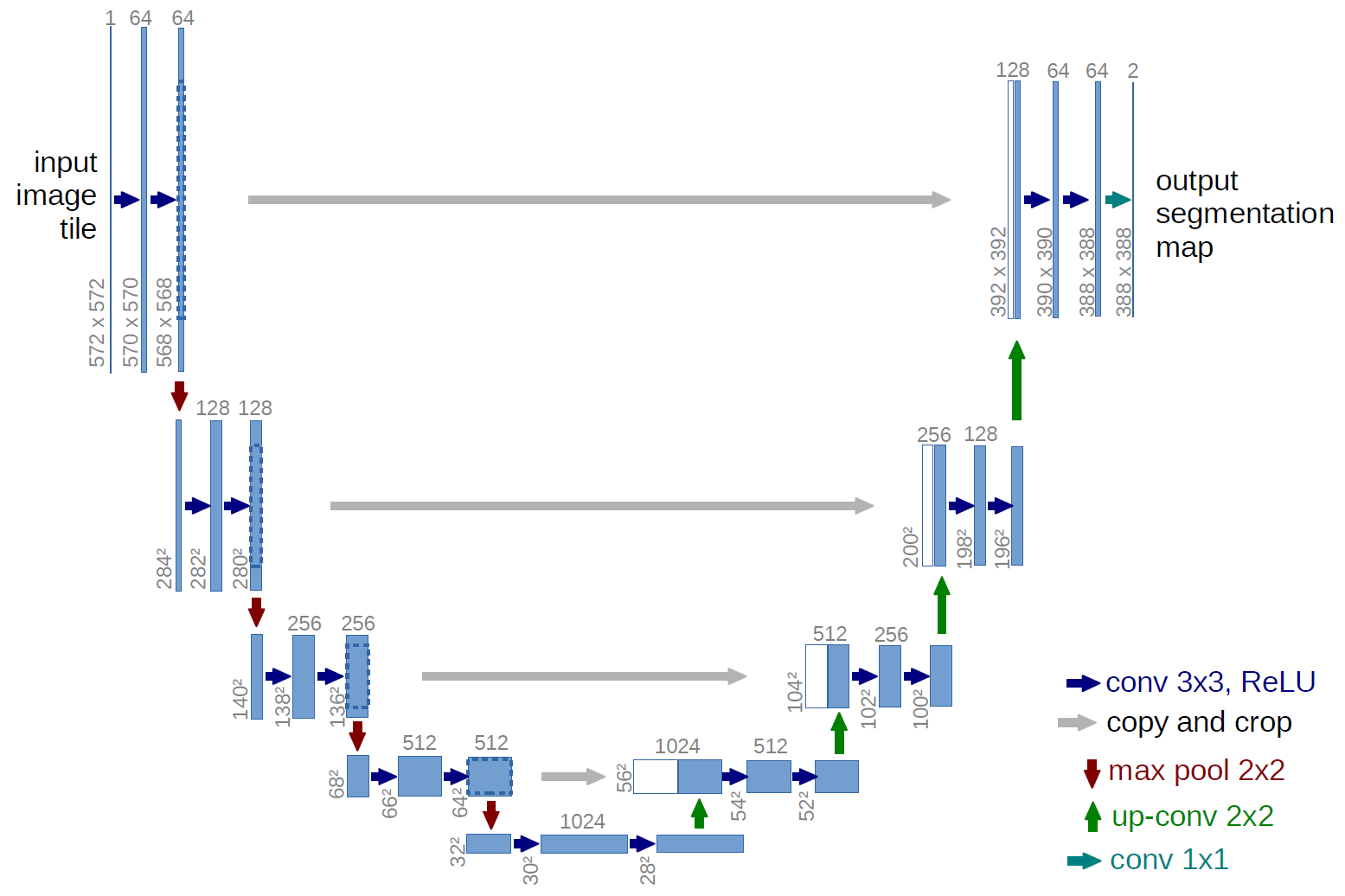}
\end{center}
\caption{\label{label}Depiction of the UNet Architecture \cite{hello}}
\end{figure*}

\begin{figure*}
\begin{center}
\includegraphics[width=30pc]{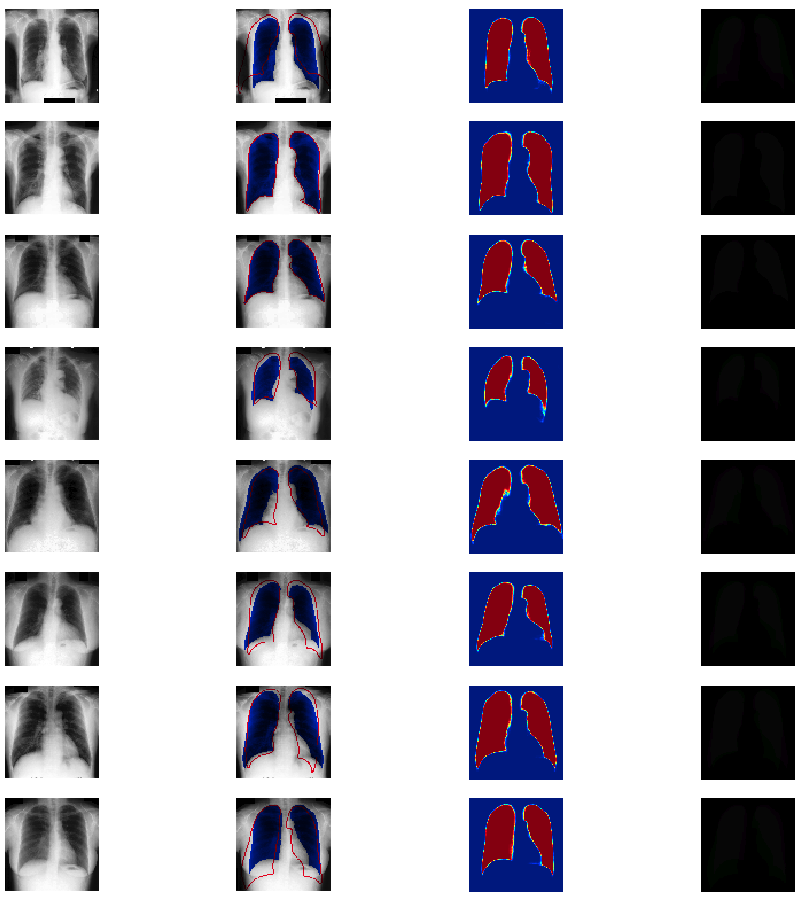}
\end{center}
\caption{\label{label}Sample of the results obtained by the proposed approach}
\end{figure*}

\section{Methodology}
\subsection{Preprocessing}
The methodology for the challenge followed a bottom-up approach, starting from basic statistical image processing algorithms, gradually building up on the findings and moving towards iteratively complex architectures capable of capturing and utilizing features in the image more effectively.
The first step though before any algorithm could be implemented was the preprocessing step in which the dataset was analyzed and converted into a form suitable for segmentation. The dataset consisted of chest X-ray images and masks corresponding to each image for the lung portions. It was determined that the masks would need to be superimposed on the chest X-Rays to create the training set and the test set would remain in its original unaltered form.  

\subsection{Approach 1: Connected Component Analysis}
The first approach consisted of converting the chest X-Ray image to grayscale, using Otsu’s Binarization \cite{snu} to obtain a binary image from the grayscale, and subsequently using the Connected Component Analysis \cite{v} algorithm to get connected regions and joining them to obtain the outline of the lungs. Connected Component Analysis is essentially a method of joining similar contoured regions in a binary image using graph theory. This ensures that same components have pixels at similar levels of intensity which leads to efficient determination of the portion of the image to be segmented. The flowchart in Figure \ref{fig21}. provides a high level overview of the algorithm for a ready reference into the methodology.

Despite showing initial promise, the approach failed due to disparate regions getting connected together resulting in completely arbitrary outlines. This also raised the issue of false positives, since artifacts might have influenced even the correct outcomes, in the absence of a proper learning framework.

\subsection{Approach 2: Watershed Algorithm}
The second approach was implementation of the Watershed Algorithm \cite{nyu, yeongleeseng}, using the thresholding created by binarization. Watershed Algorithm is extremely adept at identifying segments in an image which touch or even overlap, and thus it was expected that the darker regions would be appropriately identified by thresholding and filled up, since the Watershed Algorithm considers darker regions to be lying at the troughs and the lighter regions to be lying at the crests. The pixels in the image are considered as points of elevation in the topography, and thus the valleys are \say{\emph{flooded}} while the adjoining regions at relatively higher altitude create a natural barrier, keeping the submerged areas from combining together. This results in the formation of natural segmentation lines in the image by partitioning off areas with similar topographical features. The detailed steps in the implementation of this approach have been given in Figure \ref{fig22}. It was expected that the algorithm would  thereby successfully delineate the lungs from the X-Rays. 

Unfortunately, this approach fell short of expectations as well due to a problem with the surface on which the X-rays were kept prior to being scanned. Since the colour of the surface and the X-ray were similar, the output tended to include the surface as well, and with no method to account or correct for this problem, it was decided to discard this approach in favour of neural networks which minimize their error via back-propagation, in essence, learning from their mistakes after each output and trying to minimize it.

\begin{table*}[]
 \caption{Comparative Result Analysis}
    \label{table2}
    \begin{center}
    \begin{tabular}{|C{4cm}|C{4cm}|C{4cm}|}
        \hline
		\textbf{Name of Approach} & \bf IoU Metric & \bf DICE Score \\
		\hline
		Connected Component Analysis & 42.6 & 46.2 \\
		\hline
		Watershed Algorithm & 52.8 & 59.7 \\
		\hline
		U-Net Model & 78.4 & 82.7 \\
		\hline
    \end{tabular}
    \end{center}
\end{table*}

\subsection{Proposed Approach: UNet}
The UNet deep learning algorithm is considered to be the current state-of-the-art architecture for biomedical image segmentation, overtaking the previous Convolutional Neural Network models. This high performance is due to the innovative up-sampling technique employed in UNets which helps to segment images using the corresponding down-sampled feature mappings through skip connections. This unique innovation in the traditional CNN model gives it the name UNet and the implemented model performed exceedingly well on our dataset as well. The dataset was split in an 8:1:1 ratio for the training set, the testing set, and the validation set respectively, and 10-Fold cross validation was done to improve the robustness of the model.

The structure of the UNet architecture has been shown in Figure 3 which also makes the choice of its name clear owing to the \say{U} shape, successively up-sampling the feature extractions to preserve image quality and efficiently perform semantic segmentation \cite{septicflesh}.

\section{Comparative Results Analysis}

The performance of the three models was measured using the IoU metric and the DICE score (akin to the F1 score), as given in Table 1. An impressive DICE score of 82.7 was obtained for the U-Net model despite the small size of the dataset, establishing the efficiency of the approach. A part of the output has been given in Figure 4, where  the first column contains the preprocessed images, the second one is the superimposition of the predicted segment against the actual segmentation done by hand, the third column shows the predicted masks for each sample and the last column contains a graphical representation of the difference between the predicted value and the optimum value.

Even without an extensive quantitative evaluation, it is clear by just taking a cursory glance at the result samples that the performance of the model nearly matches the segmentation done manually by an expert. The overlap is significant and up-sampling provides a convenient way to readily analyze the results.

\section{Conclusion and Future Work}
Even though a lot of work needs to be done before the proposed lung segmentation model can be implemented in the field, it establishes a good baseline and gives ample scope for further improvements. UNet is clearly the best performing model for this task, and suitable experimentation has been carried out before arriving at that conclusion, going through various approaches and improving over the shortcomings of successive models, thus arriving at UNet in a logical fashion. 

This work is also of special relevance in the current scenario, where the COVID-19 pandemic has wreaked havoc across the world and lack of proper diagnostic resources are exacerbating the situation. Even without the pandemic, the work is important since lung disorders continue to plague healthcare systems around the world and the low-resource intensive pipeline developed here would be vital in ameliorating the condition.

Future work should revolve around training the model on a larger dataset, with augmentation techniques to improve the robustness of the model and to guard it against overfitting, while gradient boosting should be used as an integration into the final step to try and increase the learning rate in fewer number of epochs. Extension of the work into other areas, such as orthopaedics, cardiology, etc. would also be highly beneficial and provide a comprehensive diagnostic framework through highly accurate and automated tools \cite{jiaming}.

\end{document}